\documentclass[aps, prb, reprint,superscriptaddress,amsmath,amssymb]{revtex4-1}
\usepackage{graphicx}%
\usepackage{dcolumn}%
\usepackage{bm}%
\usepackage{hyperref}%
\hypersetup{backref, colorlinks=true,linkcolor=blue, urlcolor=blue, citecolor=blue,linktocpage, linktoc=all}%
\usepackage{epstopdf}%
\usepackage{bbding}%
\draft

\usepackage{color}

\begin{document}
\title{Excitons in InP, GaP, GaInP quantum dots: Insights from time-dependent density functional theory}%
\author{Xiaoyu Ma}
\affiliation{Key Laboratory for Special Functional Materials of Ministry of Education, Collaborative Innovation Center of Nano Functional Materials and Applications, and School of Materials Science and Engineering, Henan University, Kaifeng, Henan 475001, China}
\author{Jingjing Min}
\affiliation{Key Laboratory for Special Functional Materials of Ministry of Education, Collaborative Innovation Center of Nano Functional Materials and Applications, and School of Materials Science and Engineering, Henan University, Kaifeng, Henan 475001, China}
\author{Zaiping Zeng}
\email{zaiping.zeng@henu.edu.cn}
\affiliation{Key Laboratory for Special Functional Materials of Ministry of Education, Collaborative Innovation Center of Nano Functional Materials and Applications, and School of Materials Science and Engineering, Henan University, Kaifeng, Henan 475001, China}
\author{Christos S. Garoufalis}
\email{garoufal@upatras.gr}
\affiliation{Materials Science Department, University of Patras, 26504 Patras, Greece}
\author{Sotirios Baskoutas}
\affiliation{Materials Science Department, University of Patras, 26504 Patras, Greece}
\author{Yu Jia}
\email{jiayu@henu.edu.cn}
\affiliation{Key Laboratory for Special Functional Materials of Ministry of Education, Collaborative Innovation Center of Nano Functional Materials and Applications, and School of Materials Science and Engineering, Henan University, Kaifeng, Henan 475001, China}
\affiliation{International Laboratory for Quantum Functional Materials of Henan, and School of Physics and Engineering, Zhengzhou University, Zhengzhou, Henan 450001, China}
\author{Zuliang Du}
\affiliation{Key Laboratory for Special Functional Materials of Ministry of Education, Collaborative Innovation Center of Nano Functional Materials and Applications, and School of Materials Science and Engineering, Henan University, Kaifeng, Henan 475001, China}
\date{\today}
\begin{abstract}
Colloidal quantum dots (QDs) of group III-V are considered as promising candidates for next-generation environmentally friendly light emitting devices, yet there appears to be only limited understanding of the underlying electronic and excitonic properties. Using large-scale density functional theory with the hybrid B3LYP functional solving the single-particle states and time-dependent density functional theory accounting for the many-body excitonic effects, we have identified the structural, electronic and excitonic optical properties of InP, GaP and GaInP QDs containing up to a thousand atoms or more. The calculated optical gap of InP QD appears in excellent agreement with available experiments, and it scales nearly linearly with the inverse diameter. The radiative exciton decay lifetime is found to increase surprisingly linearly with increasing the dot size. For GaP QDs, we predict an unusual electronic state crossover at diameter around 1.5 nm whereby the nature of the lowest unoccupied molecular orbital (LUMO) state switches its symmetry from $\Gamma_{5}$-like at larger diameter to $\Gamma_{1}$-like at smaller diameter. After the crossover, the absorption intensity of the band-edge exciton states is significantly enhanced. Finally, we find that Vegard's law holds very well for GaInP random alloyed quantum dots down to ultra-small sizes with less than a hundred atoms. The obtained energy gap bowing parameter of this common-cation compound in QD regime appears positive, size-dependent and much smaller than its bulk parentage. The volume deformation, dominating over the charge exchange and structure relaxation effects, is mainly responsible for the QD energy gap bowing. The impact of excitonic effects on the optical bowing is found to be marginal. The present work provides a road map for a variety of electronic and optical properties of colloidal QDs in group III-V that can guide spectroscopic studies.
\end{abstract}
\maketitle

\section{Introduction}
Colloidal quantum dots (QDs) have demonstrated their great potential in modern light emitting devices\cite{Mashford407, Dai96, Chen765} owing to their high stability, tunable emission spectrum, narrow bandwidth, and broad luminescent spectral range. Light emitting diode (LED) technology based on cadmium selenide (CdSe) QDs of group II-VI has witnessed tremendous development in the last two decades, with both brightness and external quantum efficiency rivaling the state-of-the-art organic light emitting devices\cite{Shen192}.  However, the heavy metal-containing feature of these group II-VI QDs is the major obstacle limiting their further development towards commercialization. Colloidal QDs of group III-V compounds (InP, GaP, GaInP, etc.) have been used for a plethora of applications, such as color converter in liquid crystal display\cite{Owen10939}, LEDs\cite{Lim9019, Jo17}, thin-film transistors\cite{Jang653}, and bioimaging\cite{Bharali11364, Bruns0056}. Among them, InP is considered as a promising candidate to replace CdSe as a material of choice for commercial QD displays due to its low toxicity\cite{Chibli2552, Lin341} but comparable, or even broader emission color range over traditional group II-VI compounds. The other members of group III-V family such as GaP and GaInP have also seen a significant surge of interest as light emitting materials\cite{Kornienko3951, Sun19306, Lavina2432, Srivastava12144}. Even though the synthetic chemistry of colloidal III-V semiconducting QDs has seen significant progress\cite{Ramasamy6893, Srivastava3623,Franke12749}, the growth of uniform, monodisperse high quality QDs of group III-V remains challenging. This is partly because of the intrinsic more covalent bonding nature of group III-V compounds, and partly due to the lack of appropriate cation and anion precursors with balanced reactivity\cite{Xu5331}. The resultant low quality of the fabricated QDs therefore hinders the exploration of their excitonic optical properties, and more broadly limits their device applications.

Modelling of the excitonic properties of a QD requires, (i) accurate ground-state electronic structure calculations, which is able to deliver correct band gap and atomistic wave functions, and (ii) accurate treatment of excited state properties. In the former aspect, atomistic tight binding method\cite{Niquet5109, Zielinski115424, Lee195318} and empirical pseudopotential theory\cite{Wang9579, Franceschetti1819, baskoutas9301, Zeng125302, Zeng235410} are able to describe the electronic properties of QDs from few hundreds atoms to millions atoms. However, those methods are heavily parameterized, and usually rely on a pre-defined unrelaxed geometry due to the lack of total energy calculations. Continuum models, such as effective-mass approximation and $\textbf{k} \cdot \textbf{p}$ theory, are best suit for large QDs, but fail where atomistic effects become important. On the latter aspect, many-body effects play an important role in the excitonic optical properties, such as absorption edge, excitonic polarization, and fine structure. This interaction is largely magnified in a zero-dimensional system due to the combined effects of geometric confinement and reduced screening.

In this work, we study theoretically the electronic and excitonic optical properties of colloidal QDs of typical group III-V compounds, such as InP, GaP and GaInP, employing the ground-state and excited state density functional theory calculations (i.e., DFT and TDDFT). Thanks to the group theory and high-performance computing facilities, we are able to treat realistic QDs with more than 1000 atoms. We have determined a variety of excitonic optical properties of those QDs, including size-dependent optical gap, exciton binding energy, exciton decay lifetime, singlet-triplet splitting and optical absorption spectrum. In the following section, we will outline the computational details. Thereafter, in Sec. \ref{sec_rd}, numerical results and related discussion are presented. Sec. \ref{sec_col} is devoted to conclusions.

\begin{figure}[t!]
\centering
\includegraphics[scale=0.89]{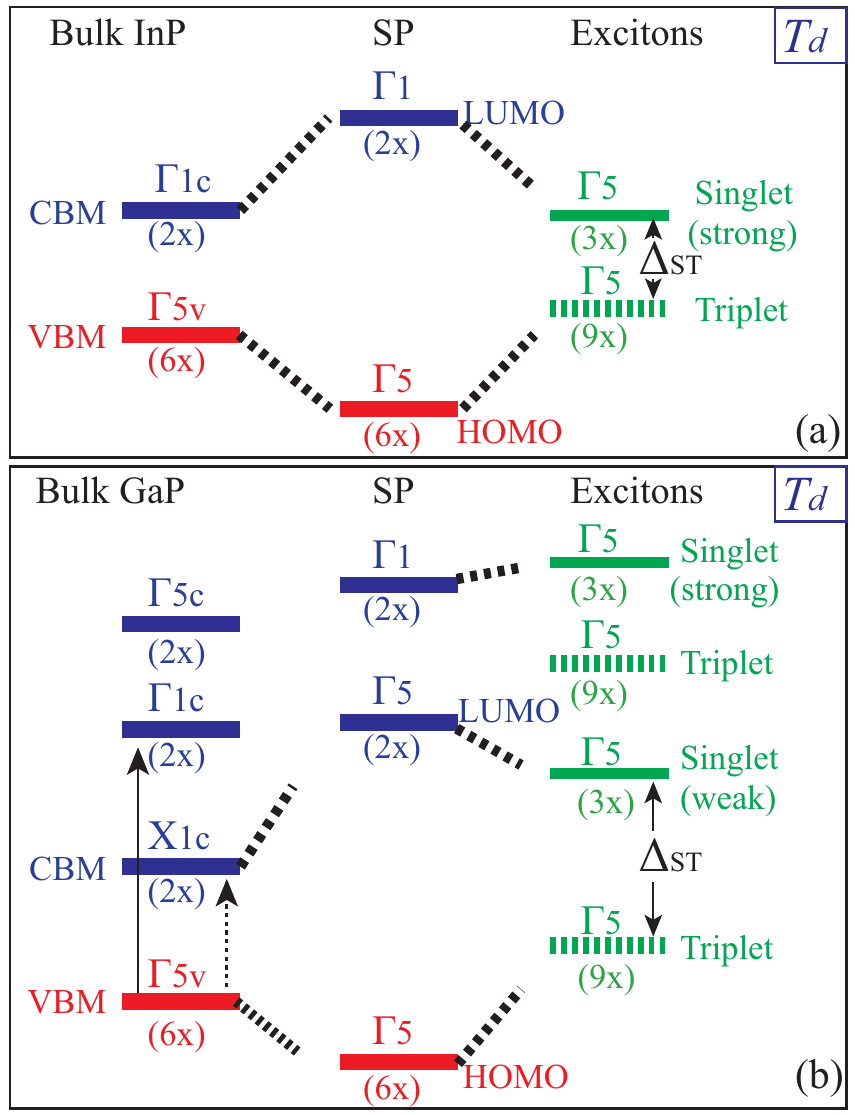}%
\caption{\label{fig1} (a) Symmetry characters of valance band maximum (VBM) and few topmost conduction band states in bulk (a, left column) InP and (b, leftmost column) GaP, the HOMO and first few LUMO states of the corresponding quantum dots at the single-particle (SP) level obtained by DFT/B3LYP method (central column), and the resulted exciton manifolds obtained from TDDFT calculations (rightmost column). The degeneracy of the energy levels or exciton states is shown in the parenthesis. In the leftmost column, the solid vertical arrow indicates an optically allowed transition, while the dashed vertical arrow indicates an optical forbidden one. In the rightmost column, a thick solid horizontal line indicates a symmetry allowed and spin allowed exciton state (singlet state), and a thick dashed horizontal line indicates a symmetry allowed by spin-forbidden exciton state (triplet-state). The spin-orbit interaction is neglected but the exchange interaction is considered. We have employed Kosters notations of single group symmetry representations within $T_{d}$ point group.}
\end{figure}

\section{Computational details}

The quantum dots (QDs) are cut from the corresponding bulk materials with nearly spherical shape (i.e., characterized by diameter $D$) and centered at a cation atom. This naturally leads to a $T_{d}$ point group symmetry for a nearly spherical QD of zinc-blende structure. The sizes of QDs range from 1.07 nm to nearly 3.5 nm which has total number of atoms ranging from 65 atoms to 1101 atoms, and therefore are within the strong confinement regime. QDs of such sizes can be synthesized using the well-established modern colloidal fabrication method\cite{Micic4904, Micic95, Cho215201}. The surface dangling bonds are passivated with pseudohydrogens, which has modified nuclear charges of 1.25 and 0.75 to terminate surface cations and anions of group III-V (e.g., InP, GaP \emph{et al.}), and of 1.5 and 0.5 to terminate their counterparts of group II-VI (e.g., CdSe), respectively. Those pseudohydrogen atoms have been chosen as the simplest model ligand, and is expected to well reproduce the size-dependent experimental band gaps of various colloidal QDs.

\begin{figure*}[t!]
\centering
\includegraphics[scale=0.65]{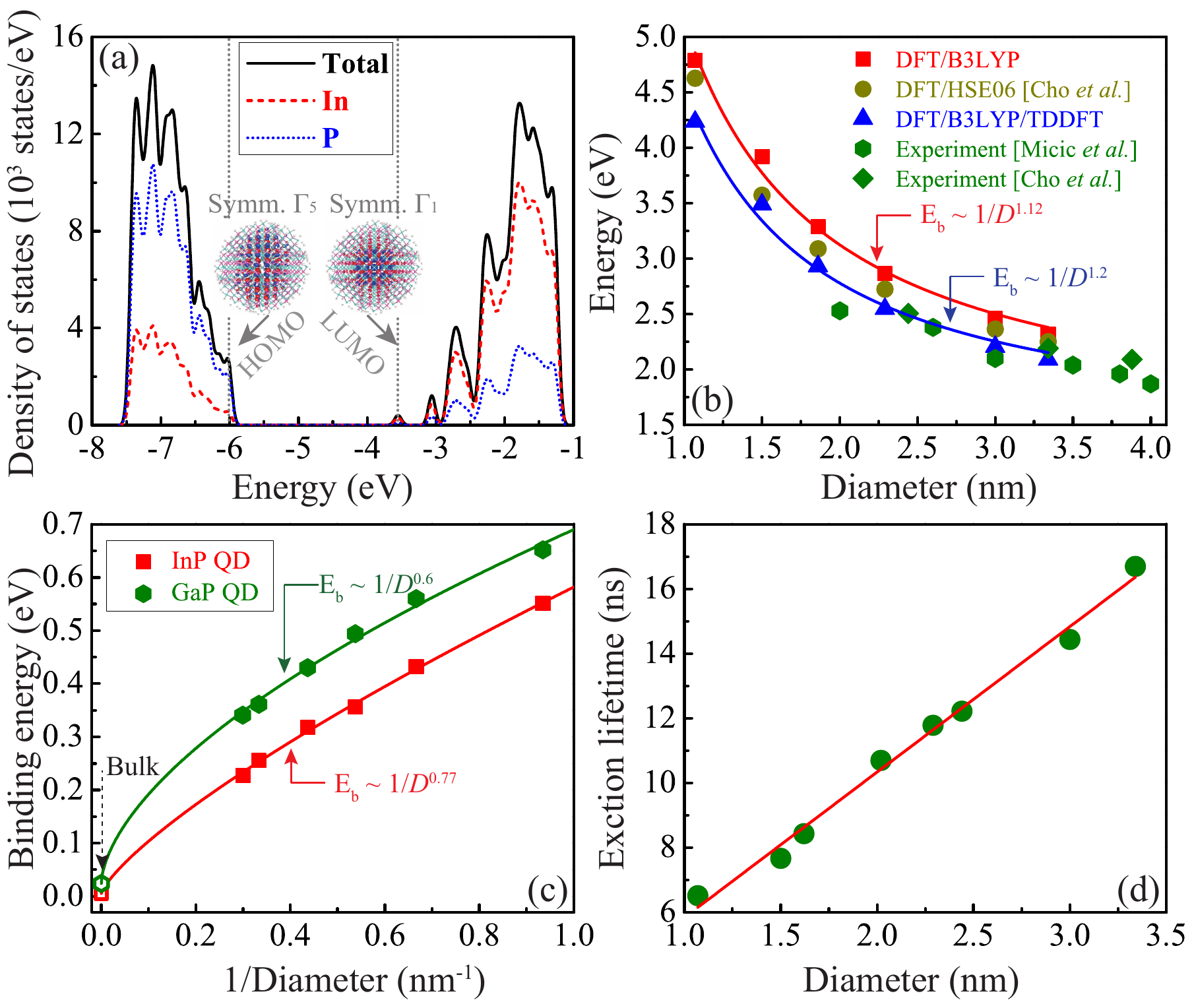}%
\caption{\label{fig2} (a) Atom resolved density of states of InP quantum dots with diameter $D=3$ nm. (b) Energy gap as a function of the diameter of InP quantum dots calculated using DFT/B3LYP and DFT/B3LYP/TDDFT, respectively, compared with existing theories and experiments\cite{Micic4904, Cho215201}. Each solid line represents a fit according to Eq. (\ref{eq2}). (c) Exciton binding energies as a function of the diameter of InP and GaP quantum dots. (d) Exciton decay lifetime as a function of the diameter of InP quantum dots. The solid line represents a linear fit.}
\end{figure*}

All calculations are performed with the Turbolmole suit of programs\cite{TURBOMOLE}. The geometry optimization is performed in the framework of density functional theory (DFT) with generalized gradient approximation (GGA) of Perdew-Burke-Ernzerhof (PBE) type\cite{Perdew3865}, which is known to predict rather accurately the structural properties. However, this level of theory is known to underestimate the band gap, and is detrimental for the modelling of optical properties. We therefore employ the hybrid nonlocal exchange-correlation functional of Becke and Lee, Yang and Parr (B3LYP \cite{Stephens11623}) to calculate reliably the single-particle HOMO-LUMO gap. We have chosen a basis set of double zeta quality (namely, the def2-SVP basis sets of the Karlsruhe group\cite{Weigend3297, Weigend1057}) throughout the work, which allows for calculations in systems with tens of hundreds of atoms without significantly compromising the accuracy.

The excitonic optical properties are calculated on top of the B3LYP results, using the linear-response time-dependent DFT (TDDFT). We note that the energy of the lowest symmetry-allowed and spin allowed transition (singlet state) is considered as the optical gap regardless of its oscillator strength. The radiative decay lifetime ($\tau_{X}$) is calculated according to\cite{Dexter1958, Zhang075404}, $\frac{1}{\tau_{X}} = \frac{4 \alpha E_{X} n |M_{X}|^{2}}{m^{2}_{0} \hbar c^{2}}$, where $n$ is the refractive index, $\alpha$ is the fine-structure constant, $m_{0}$ is the electron rest mass, and $c$ is the velocity of light, $E_{X}$ is the exciton energy, and $M_{X}$ is the electric dipole moment obtained from the TDDFT calculations. The singlet-triplet splitting is defined as the energy difference between the lowest singlet and triplet states based on the optimized ground-state geometry.

\section{Results and discussion}
\label{sec_rd}
\subsection{InP quantum dots}

Bulk InP is known to be a direct gap semiconductor with a gap of 1.4236 eV at cryogenic temperature\cite{Mathieu4042}. Bulk VBM is six-fold (without spin-orbit interaction) and of $\Gamma_{5v}$ symmetry, while the CBM is two-fold and of $\Gamma_{1c}$ symmetry (cf. Fig. \ref{fig1}(a)). The optimized In-P bond lengths of the interior atoms in the InP QDs at GGA/PBE level of theory are ranging from 2.591 $\rm \AA$ to 2.594 $\rm \AA$, which are nearly identical to that of bulk InP structure ($\sim$ 2.598 $\rm \AA$). The charge density distribution of the highest occupied molecular orbital (HOMO) state of InP QDs mainly resides on P atoms, having a $p$-type character and a $\Gamma_{5}$-like symmetry similar to its bulk parentage (cf. Fig. \ref{fig2}(a)). The lowest unoccupied molecular orbital (LUMO) state is contributed from the $s$-orbitals of both In and P atoms (cf. Fig. \ref{fig2}(a)), and the corresponding charge density distribution accumulates around both types of atoms (cf. Fig. \ref{fig2}(a)). It has a $\Gamma_{1}$ symmetry, having the same symmetry as the bulk CBM (cf. Fig. \ref{fig1}(a)). The $s$-$p$ coupling in a bonding/anti-bonding manner leads to the opening of the electronic band gap of InP QDs. The HOMO state lowers down in energy, and the LUMO state rises up in energy when enhancing quantum confinement effects, as expected.  We find that comparing to the HOMO state, its LUMO counterpart varies more significantly in energy, therefore suggesting that the electron is more delocalized and more sensitive to the quantum confinement effects.

We next evaluate the size-dependent band gap of InP QDs. It is known that the size-dependent band gap ($E_{g}$) of a QD is simply expressed according to the analytical equation\cite{Brus4403, Nanda605},
\begin{equation}
\label{eq2}
E_{g}=E_{g, bulk} + C_{g}/D^{\alpha},
\end{equation}
where $E_{g, bulk}$ is the bulk band gap, $C_{g}$ is a proportionality constant, and $\alpha$ is a real number. We study the band gap evolution as a function of the dot size in Fig. \ref{fig2}(b) based on two levels of theory, i.e., single-particle level using DFT with the B3LYP hybrid functional and correlated exciton level employing TDDFT. The exciton effects are neglected at the former level of theory, and they are properly accounted for at the latter level of theory. A side-by-side comparison of the results at both levels of theory allows us to quantify the impact of excitonic effects on the scaling law. We find that the optical gap is systematically lower than the single particle gap giving a physically consistent picture. This is not always true in DFT calculations, since it is quite common for pure GGA functional to give TDDFT optical gaps larger than the corresponding single-particle ones. Moreover, the calculated optical gap values compare well with the available experiments (cf. Fig. \ref{fig2}(b)). The calculated gaps as a function of the dot size is well fitted with the analytical equation (\ref{eq2}), delivering $C_{g}=1.12$ at single-particle level. The excitonic effects turn out to have marginal impact on the scaling law, increasing $C_{g}$ slightly to 1.2. Those values obtained from the current \emph{ab \ initio} method differ significantly from that predicted from single-band effective-mass theory using the particle-in-box model ($C_{g} = 2$), and from that obtained based on large-scale atomistic empirical pseudopotential theory $C_{g} = 1.36$)\cite{Fu1496}. However, such a nearly linear scaling of energy gaps with respect to the inverse diameter agrees well with the experimental measurements on the core/shell QDs ($C_{g} \approx 1$)\cite{Narayanaswamy2539}. We note that the actual quantum confinement effects of InP QDs shall be different from that of InP/ZnS core/shell QDs in two ways. Firstly, the structure properties of InP QDs can be significantly affected by the shell. The interfacial strain, induced by the lattice mismatch between the core and shell, can reach up to 4\%, depending on the core size and the shell thickness\cite{Park249}. Secondly, the screening effects of core and that of the shell are very different, which shall have profound effects on the optical properties.

\begin{figure*}[t!]
\centering
\includegraphics[scale=0.7]{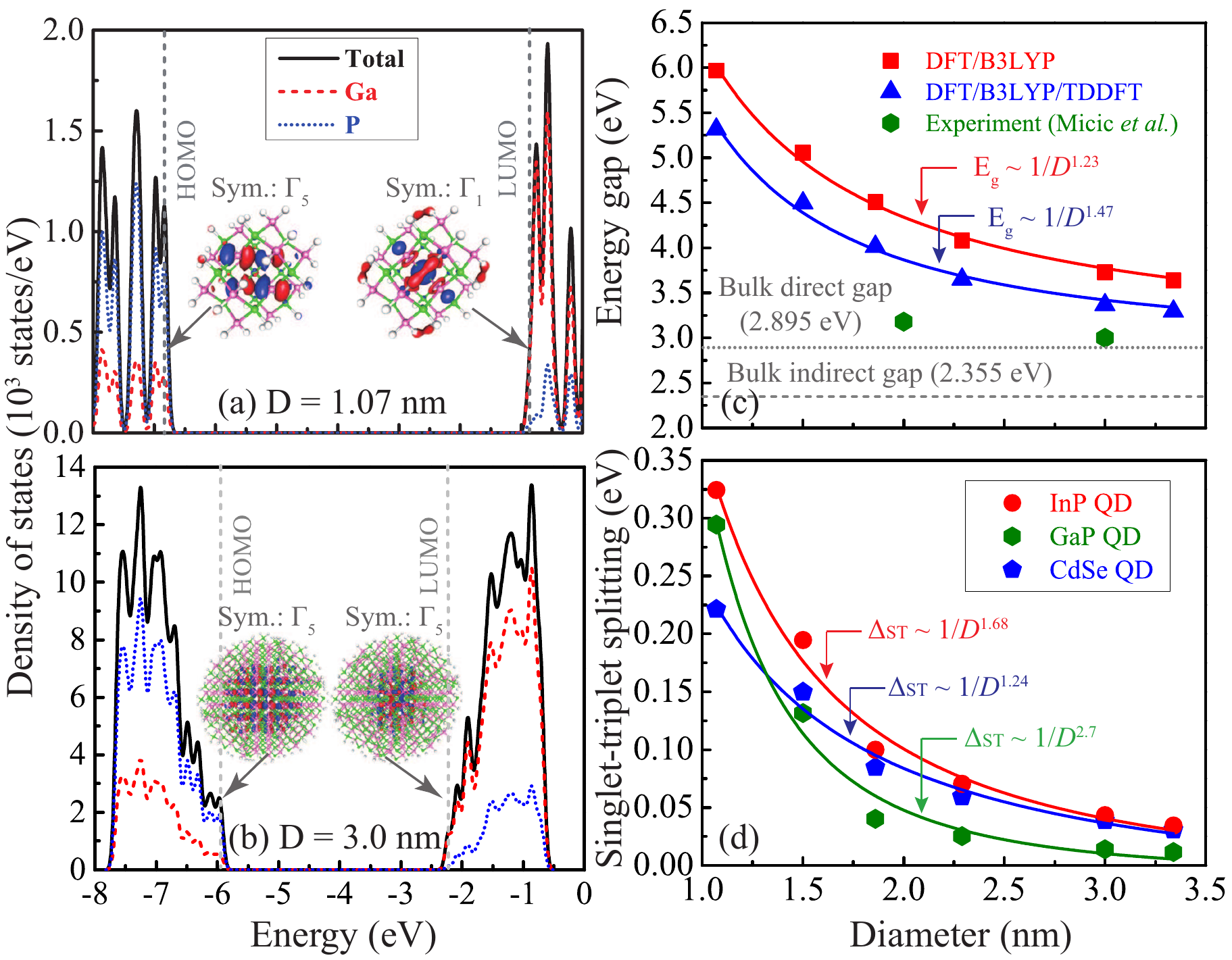}%
\caption{\label{fig3} Atom resolved density of states of GaP quantum dots with diameter (a) $D=1.07$ nm and (b) 3 nm obtained at DFT/B3LYP level of theory, respectively. The dotted vertical line indicates the energy position of HOMO and LUMO states. The inset shows the charge density of the HOMO and LUMO states alongside with the corresponding symmetry character. (c) Energy gap as a function of diameter of GaP quantum dots calculated using DFT/B3LYP and DFT/B3LYP/TDDFT levels of theory, respectively, compared with existing experiments\cite{Micic95}. Each solid line represents a fit according to Eq. (\ref{eq2}). (d) The singlet-triplet splitting $\Delta_{\rm ST}$ as a function of the diameter of GaP QDs, compared with that of InP and CdSe QDs. Each solid line represents a fit using equation, $\Delta_{\rm ST}= \delta+C_{st}/D^{\gamma}$, where $\delta$, $C_{st}$ and $\gamma$ are fitting parameters.}
\end{figure*}

In contrast to the multiplicity of both experimental and theoretical activities on physical properties in ``high energy" ($\sim$ 1 eV) scale, the focus on the properties of InP QDs in ``low energy" ($\sim$ 10$^{-3}$-eV) scale is rather limited. We therefore study the size-dependence of exciton binding energy in InP QDs, which appears in the ``low energy" scale (cf. Fig. \ref{fig2}(c)). This quantity is defined as the energy difference between the single-particle gap and the optical gap. In contrast to the weakly binding nature of exciton in bulk InP ($E^{X}_{b, bulk} = 5.1$ meV\cite{Mathieu4042}), exciton under three-dimensional quantum confinement appears strongly bound, reaching up to 550 meV for our smallest InP QD. The calculated exciton binding energy ($E^{X}_{b}$) can be well fitted using the analytical equation similar to equation (\ref{eq2}),
\begin{equation}
\label{eq3}
E^{X}_{b}=E^{X}_{b, bulk} + C^{X}_{b}/D^{\beta},
\end{equation}
with $\beta$ being a fitting parameter. We find that the exciton binding energy scales as $D^{-0.77}$ (cf. Fig. \ref{fig2}(c)), which again differs from the prediction of effective-mass theory using the particle-in-box model ($E^{X}_{b} \propto \frac{1}{D}$), representing solely the Coulomb interaction between the electron and hole. For comparison purpose, we have also examined the scaling law of size-dependent exciton binding energy for CdSe QDs on an equal footing. It turns out that the calculated exciton binding energy of CdSe QD is very comparable to its InP counterpart of equal size, and it scales as $E^{X}_{b} \propto 1/D^{0.72}$, which reproduces exactly the experimentally determined scaling law\cite{Meulenberg325}. We note that the numerically obtained exciton binding energies are significantly smaller than the experiments\cite{Meulenberg325} (not shown), suggesting that the current theoretical scheme may underestimate the exciton binding energy.

Finally, we study the size-dependence of exciton decay lifetime of the first-bright exciton state of InP QDs in Fig. \ref{fig2}(d). The obtained lifetime appears at the nanosecond time scale for the sizes considered herein. Strikingly, we find that the lifetime increases monotonically, and scales linearly as a function of the dot size (cf. Fig. \ref{fig2}(d)). Such a linear behavior has also been experimentally found for CdSe QDs\cite{Gao4230}. We find that the lifetime of InP QD appears systematically larger than its CdSe counterpart of equal size and stoichiometry. We note that the lifetime not only depends on the dot size, but also on the surface inorganic stoichiometry.

\subsection{GaP quantum dots}

Gallium phosphide (GaP), as another typical compound of group III-V, is known to be an indirect band gap semiconductor with VBM locating at $\Gamma$-point and of $\Gamma_{5}$ symmetry, and CBM locating at $\rm X$ point and of $\rm X_{1}$ symmetry (cf. Fig. \ref{fig1}(b)). It has an indirect gap of 2.355 eV\cite{Humphreys5590} and a larger direct gap of 2.895 eV\cite{NelsonA1399} at $\Gamma$-point measured at cryogenic temperature. GaP therefore has been considered as a promising candidate for a blue fluorophore. However, given its indirect band gap nature, GaP can not be considered as an efficient photon emitter in bulk at least at cryogenic temperature where phonon-assisted emission is minimal. One way to make GaP relevant as a light-emitting compound is to alleviate the indirect nature of the band structure via quantum confinement effects. Many experimental activities therefore have focused on synthesis of GaP QDs under strong quantum confinement regime\cite{Micic95, Kim13656, Kher2056, Gao3064, Manciu4059, Kim2466}. Both broad size distribution and the lack of thorough analysis of the optical properties of the fabricated GaP QDs result in only a rough determination of the corresponding quantum confinement effects. The reported \emph{ab initio} study has been limited to cluster size\cite{Kamal024308}. We therefore employ the aforementioned reliable theoretical scheme successfully applied on the study of InP QDs to gain insights on the excitonic optical properties of GaP QDs of realistic sizes.

The optimized Ga-P bond lengths of the interior atoms in the QDs are nearly identical to their bulk parentage ($\sim$ 2.3945 $\rm \AA$), regardless of the dot size. The HOMO state of GaP QD always has a dominant contribution from $p$-orbital of P atoms, and inherits the same symmetry character of the corresponding bulk phase (e.g., $\Gamma_{5}$, cf. Fig. \ref{fig1}(b), Fig. \ref{fig3}(a) and (b)). In contrast to bulk CBM having a $X_{1}$ symmetry, the LUMO state of GaP QD presents a $\rm \Gamma_{5}$ symmetry for larger QDs (e.g., $D=3$ nm and cf. Fig. \ref{fig3}(b)), which switches to $\rm \Gamma_{1}$ symmetry at ultra-small sizes (e.g., $D=1.07$ nm, cf. Fig. \ref{fig3}(a)), therefore recovering the symmetry of bulk CBM at $\Gamma$ point (cf. Fig. \ref{fig1}(b)). Such a switching is found to occur at diameter around 1.5 nm. The single-particle gap of GaP QD is calculated in Fig. \ref{fig3}(c) as a function of the QD size. It shows that the HOMO-LUMO gap scales as $E_{g} \propto 1/D^{1.23}$, indicating that GaP QD is more sensitive to the quantum confinement effects than its InP counterpart (e.g., $E_{g} \propto 1/D^{1.12}$).

The direct transition from HOMO to LUMO states results in a nine-fold spin-triplet states (optically dark) at a lower energy, and three-fold spin-singlet state (optically active) at a higher energy (cf. Fig. \ref{fig1}(b)). All of these exciton states are of $\Gamma_{5}$ symmetry (cf. Fig. \ref{fig1}(b)). Although being optically allowed, this spin-singlet exciton state exhibits weak transition dipole moment $\mu_{D}$, and therefore small oscillator strength $f_{osc}$ and long radiative decay time $\tau_{X}$. This is in stark contrast to that in InP QD. For $D=1.5$ nm, we find that $\mu^{\rm GaP}_{D} = 0.022$ Debye, $f^{\rm GaP}_{osc}=0.00016$, and $\tau^{\rm GaP}_{X}=5.86 \ \rm \mu s$. However, for InP QD of equal size, we find that $\mu^{\rm InP}_{D} = 1.54$ Debye, $f^{\rm InP}_{osc}=0.56$, and $\tau^{\rm InP}_{X}=7.66 \ \rm  ns$. Concerning the scaling law, we find that the excitonic effects bring the scaling law further apart from linear scaling. The optical gap of GaP QD is found to scale as $E^{X}_{g} \propto 1/D^{1.47}$ (cf. Fig. \ref{fig3}(c)). The calculated optical gap appears significantly larger than the reported experimental data\cite{Micic95} measured at high temperatures ($\approx 400^{\circ}$C, cf. Fig. \ref{fig3}(c)). The discrepancy shall be partly resolved by future measurements on GaP QDs with narrow size distribution at cryogenic temperatures to exclude both the size uncertainty and the thermal noise.

\begin{figure*}[t!]
\centering
\includegraphics[scale=16]{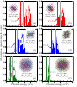}%
\caption{\label{fig4} Absorption spectrum of (a, c, e) InP and (b, d, f) GaP quantum dots with diameter (a, b) $D=1.07$ nm, (c, d) 1.5 nm and (e, f) 2.29 nm, respectively, computed on ground of 30 optically allowed exciton states. The vertical line shows the absorption peak corresponding to each exciton state. A Lorentzian broadening function is employed with broadening parameter $\Gamma_{l}=0.05$ eV. The vertical dashed line indicate the single-particle HOMO-LUMO gap, and the vertical arrows show the optical gap. The insets show the transition density corresponding to the first optically allowed exciton state.}
\end{figure*}

\begin{figure*}[t!]
\centering
\includegraphics[scale=0.65]{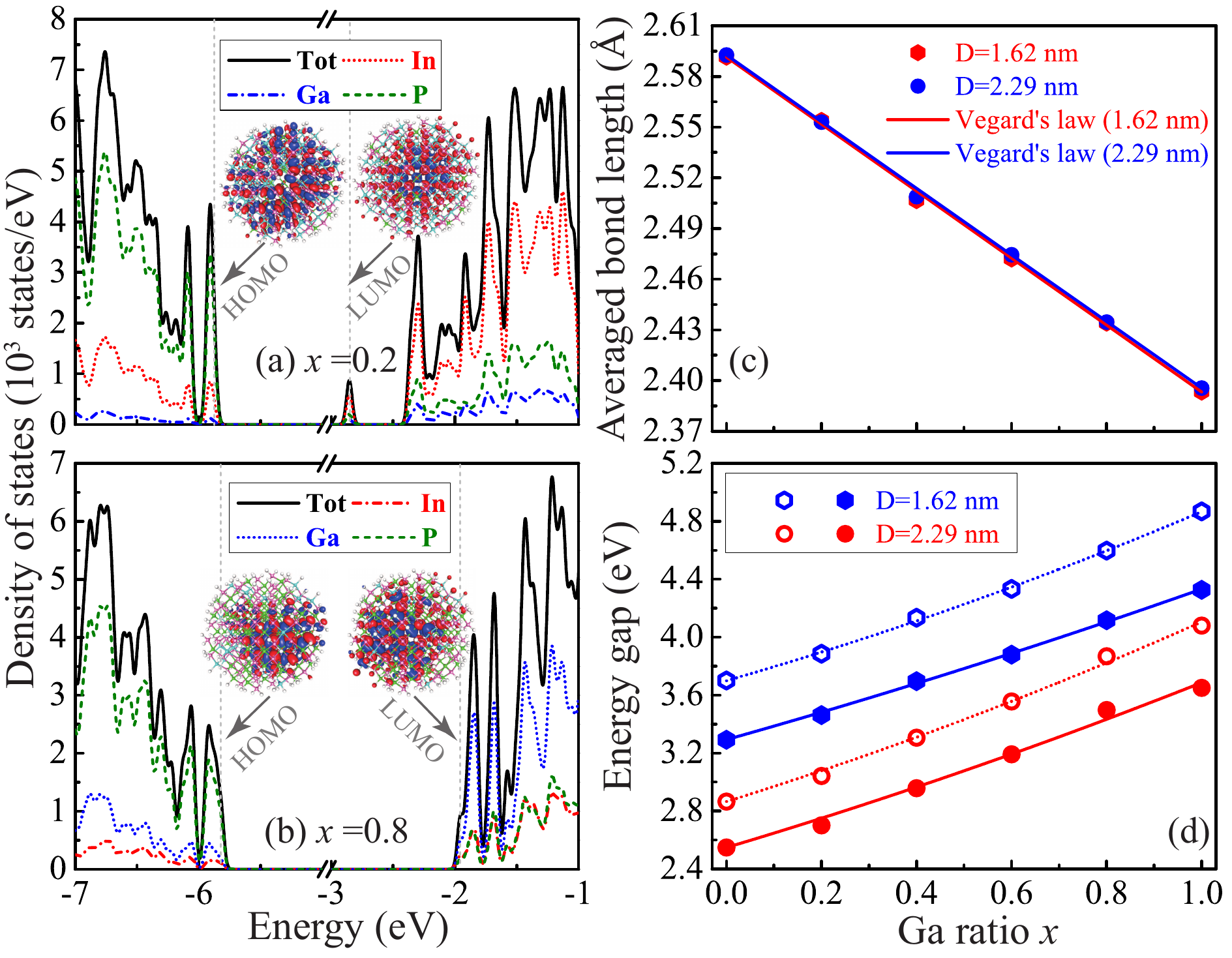}%
\caption{\label{fig5} Density of states of Ga$_{x}$In$_{1-x}$P alloyed quantum dots with Ga ratio (a) $x=0.2$ and (b) 0.6, respectively. The insets show the charge density plots of the HOMO and LUMO states. (c) The averaged bond length (symbols) of GaInP alloyed quantum dot as a function of Ga ratio $x$ for diameter $D=1.62$ and 2.29 nm, respectively. The solid line represents the expected bond length predicted from Vegard's law using the GGA/PBE calculated average bond lengths of InP and GaP QDs. (d) The single-particle (open symbols) and optical (closed symbols) gaps as a function of Ga ratio $x$ for diameter $D=1.62$ and 2.29 nm, respectively. The line represents fits according to $E_{g}=E^{\rm InP}_{g}+ax+bx^{2}$, where $a$ and the bowing parameter $b$ are fitting parameters. In (c) and (d), each data represents the averaged value over ten random geometric configurations.}
\end{figure*}

The exciton binding energy of GaP QD is systematically larger than its InP counterpart. It scales as $E^{X}_{b} \propto 1/D^{0.6}$ (cf. Fig. \ref{fig2}(c)), deviating significantly from the linear scaling law governed solely by the Coulomb interaction between the electron and hole. The singlet-triplet splitting $\Delta_{\rm ST}$ as a function of the diameter of GaP QDs is studied in Fig. \ref{fig3}(d). The results are compared with that of InP and CdSe QDs. We find that (i) $\Gamma_{5} \rightarrow \Gamma_{5}$ transition resulted exciton manifold appears to have a much smaller $\Delta_{\rm ST}$ than that originated from $\Gamma_{5} \rightarrow \Gamma_{1}$ transition. This is evident by comparing the results before ($D>1.5$ nm) and after ($D<1.5$ nm) the $\Gamma_{5}$-$\Gamma_{1}$ electronic state crossover in the LUMO state; (ii) Although both band-edge exciton manifold stemming from $\Gamma_{5}$-$\Gamma_{1}$ HOMO-LUMO transition, $\Delta_{\rm ST}$ in InP QD is systematically larger than that in CdSe counterpart. This is particularly pronounced for smaller diameters. (iii) $\Delta_{\rm ST} \propto 1/D^{1.68}$ for InP QD and $\propto 1/D^{2.7}$ for GaP QD. This indicates that $\Delta_{\rm ST}$ in QDs of group III-V is more size-dependent than their group II-VI counterpart (e.g., $\Delta_{\rm ST} \propto 1/D^{1.24}$ for CdSe QD).

The absorption spectrum of GaP QDs with various sizes is plotted and compared with that of InP QDs in Fig. \ref{fig4}. Three characteristics are observed: (i) For both types of QDs, the absorption edge blueshifts with enhancing the quantum confinement effects, as expected; (ii) GaP QDs with all sizes exhibit weaker absorption intensity than its InP counterparts, at least at lower energy part of the absorption spectrum; (iii) The first pronounced exciton absorption peak in both GaP and InP QDs are dominantly contributed from the $\Gamma_{5} \rightarrow \Gamma_{1}$ transition (cf. Fig. \ref{fig4}). For InP QD, such a transition corresponds to the HOMO-LUMO transition, and its intensity increases with \emph{increasing} the dot size due to the enhancement of overlapping between the HOMO and LUMO wave functions (cf. Fig. \ref{fig4}(a, c, e)). For GaP QDs, however, the electron state with $\Gamma_{1}$ symmetry and involved in this transition rises up, going from LUMO+5 at $D=2.29$ nm to LUMO+2 at $D=1.07$ nm. In contrast to InP QD, the peak intensity corresponds to such a transition in GaP QDs increases with \emph{decreasing} the dot size (cf. Fig. \ref{fig4}(b, d, f)). This remains true for the lower energy part of the absorption spectrum. These results therefore suggest that increasing the quantum confinements can server as an effective way of enhancing the absorption or PL intensity of GaP QDs.

\subsection{Ga$_{x}$In$_{1-x}$P random alloyed quantum dots}

After examining the electronic and excitonic optical properties of both InP and GaP QDs, we finally turn to their native ternary alloy, Ga$_{x}$In$_{1-x}$P, QDs, which is of technological importance due to the flexibility they offer in terms of band gap and lattice constant engineering. Ga$_{x}$In$_{1-x}$P QDs emitting green light are expected to be superior to their InP counterparts due to their larger size and correspondingly larger absorption cross sections and smaller surface to volume ratio. Moreover, incorporation of Ga into InP lattice reduces the lattice mismatch with wider band gap shell materials such as ZnS, making the material less strained. This will be beneficial for reducing the trap centers and slowing down Auger recombination rates. Luminescent Ga$_{x}$In$_{1-x}$P QDs with controlled composition can be well synthesized with colloidal chemistry method either in molten salts\cite{Srivastava12144} or by suitably choosing gallium precursor\cite{Wegner1663}.

\begin{figure*}[t!]
\centering
\includegraphics[scale=0.65]{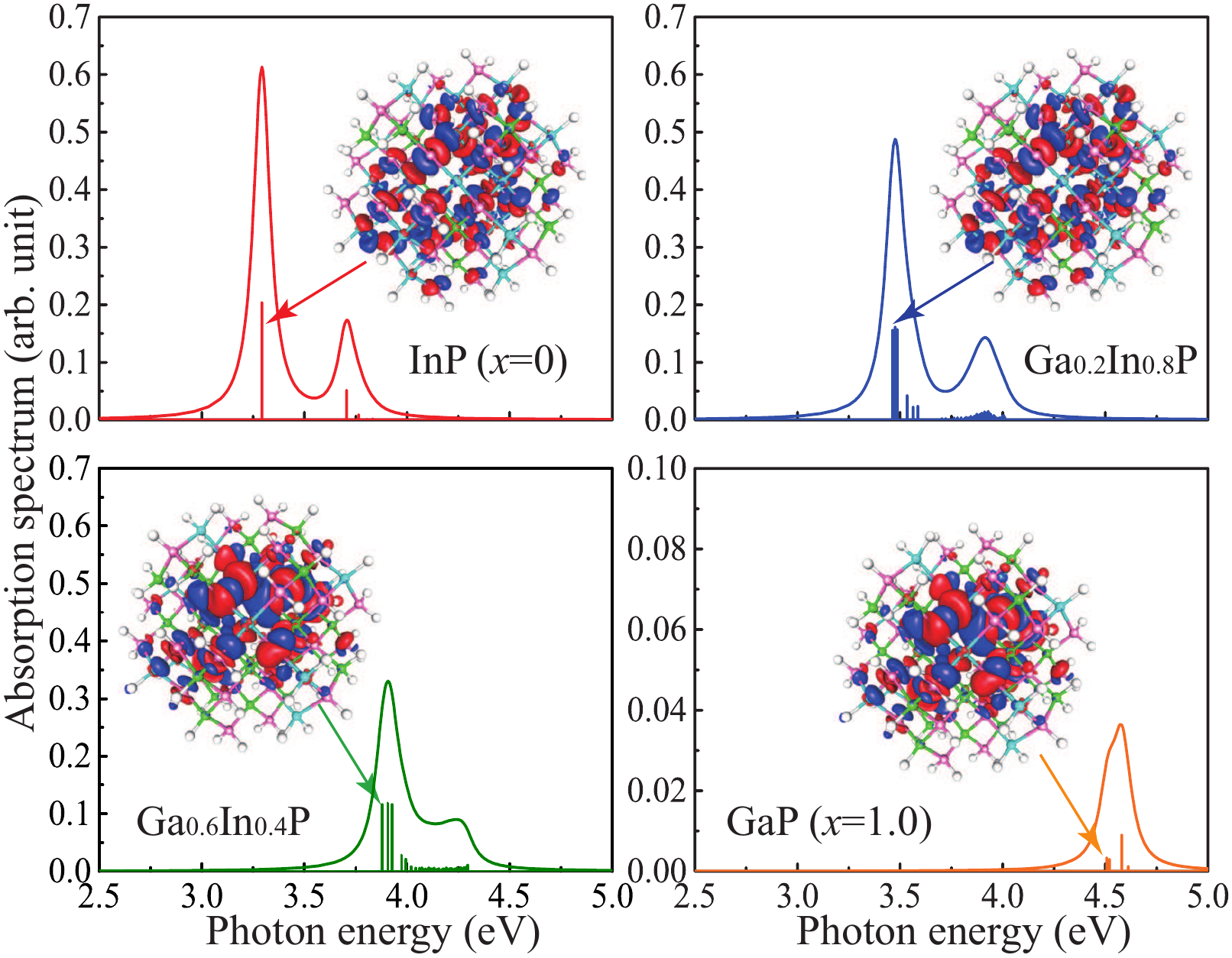}%
\caption{\label{fig6} Absorption spectrum of Ga$_{x}$In$_{x}$P quantum dots with various Ga ratio $x$ calculated with the lowest 40 exciton states without any symmetry constrains (e.g., $C_{1}$ symmetry). The vertical line shows the absorption peak corresponding to each exciton state. A Lorentzian broadening function is employed with broadening parameter $\Gamma_{l}=0.05$ eV. The inset shows the transition density corresponding to the first optically allowed exciton state.}
\end{figure*}

The lattice constant of bulk ternary compounds (A$_{x}$B$_{1-x}$C) usually varies linearly with composition $x$, e.g., $a_{\rm alloy} = x a_{AC} + (1-x)a_{BC}$. This is the well-known Vegard's law, that is empirical and based purely on observations, but hold surprisingly well for most of the bulk alloyed materials. We therefore firstly check the validity of Vergard's law when moving from bulk to the nano regime. We have chosen two representative sizes, e.g., $D=1.62$ and 2.29 nm, with the consideration of computational cost caused by the random geometric configuration averaging. We have employed GGA/PBE level of theory for the geometry optimization of the alloyed QDs, which often delivers good structural properties. Since the definition of lattice constant is no longer valid in QDs, we therefore tend to calculate the average bond length for all the fully neighboured dot atoms (e.g., excluding the surface dot atoms and the pseudohydrogen atoms), and the results are shown in Fig. \ref{fig5}(c). It is shown that Vegard's law holds well even for those ultra-small alloyed QDs, irrespectively of dot size. With varying Ga ratio $x$, we find that the HOMO state keeps its origin (e.g., $p$-orbital of P atoms, cf. Fig. \ref{fig5}(a, b)), while the LUMO switches from a dominant contribution from $s$-orbital of In atoms at smaller $x$ to a mixed contribution from $s$-orbital of both In and Ga atoms at larger $x$ (cf. Fig. \ref{fig5}(b)). The energy gap as a function of gallium ratio $x$ at both single-particle level and correlated level is shown in Fig. \ref{fig5}(d). It is found that the energy gaps at both levels experience a monotonic increase with increasing the gallium ratio. A parabola fit of the calculated data enables the determination of the linear coefficient $a$ and the bowing parameter $b$. We find that the linear coefficients at both single-particle level and correlated level are positive and close to unity, contrary to bulk band gap at $\Gamma$-point ($a^{\Gamma}_{\rm bulk}$ = -0.77 eV [Ref. \onlinecite{Merle2032}]). For example, $a=0.94$ eV at diameter $D=1.62$ nm and 1.03 eV at $D=2.29$ nm. Both values experience only a slight decrease to 0.92 eV and 0.98 eV, respectively, when considering the excitonic effects.

The obtained bowing parameters are found to be positive and size-dependent. For example, at correlated level, $b=0.16$ eV at $D=2.29$ nm and reduces to 0.116 eV at $D=1.62$ nm, both of which are much smaller than the bulk optical bowing parameter for the direct band gap at $\Gamma$-point $b^{\Gamma}_{bulk}=0.648$ eV [Ref. \onlinecite{Merle2032}]. The energy gap bowing usually can be decomposed into three physically distinct contributions\cite{Bernard5992}: (i) volume deformation $b_{vd}$, (ii) charge exchange $b_{ce}$, and (iii) structure relaxation $b_{sr}$. To identify the dominant physical contribution for the energy gap bowing of GaInP QD, we evaluate the three contributions according to\cite{Bernard5992},
\begin{eqnarray}\label{eq_bvd}
b_{vd} = \frac{E^{\rm GaP}_{g, \rm RE}-E^{\rm GaP}_{g, \rm FX}}{1-x} + \frac{E^{\rm InP}_{g, \rm RE}-E^{\rm InP}_{g, \rm FX}}{x}, \nonumber \\
b_{ce} = \frac{E^{\rm GaP}_{g, \rm FX}}{1-x}+ \frac{E^{\rm InP}_{g, \rm FX}}{x} - \frac{E^{\rm GaInP}_{g, \rm FX}}{x(1-x)}, \nonumber \\
b_{sr} = \frac{E^{\rm GaInP}_{g, \rm FX}-E^{\rm GaInP}_{g, \rm RE}}{x(1-x)},
\end{eqnarray}
where $E^{\rm InP}_{g}$, $E^{\rm GaP}_{g}$ and $E^{\rm GaInP}_{g}$ are the energy gap of InP, GaP and Ga$_{x}$In$_{1-x}$P at fully relaxed (subscript 'RE') geometry or fixed (subscript 'FX') geometry. We note that for the fixed geometry, the initial dot structure is cut from the corresponding bulk material with desired lattice constant $a_{\rm {Ga_{x}In_{1-x}P}}$ determined by Vegard's law, and then a geometry optimization procedure is applied with the dot atoms being fixed and surface passivating atoms being allowed to be fully relaxed. The energy band gap of the alloyed QD represents the averaged value over ten random geometric configurations. We find that for $D=1.62$ nm, $b_{vd}=0.51$ eV, $b_{ce}=0.12$ eV, and $b_{sr}=-0.33$ eV, respectively, at $x=0.2$. The sum of these three contribution, $b=b_{vd}+b_{ce}+b_{sr}=0.3$ eV, well reproduces the numerically obtained bowing parameter (parabolic fit of Fig. \ref{fig5}(d), $\sim$ 0.22 eV). This therefore suggests that the volume deformation is the dominant contribution responsible for the energy gap bowing of GaInP alloyed QDs. The volume deformation potential at QD regime has been found to be size-dependent and significantly reduced comparing to bulk\cite{Zeng125302, Zeng114305}. This therefore explains why the bowing parameter of GaInP QD is size-dependent and significantly smaller than its bulk parentage.

The absorption spectrum of GaInP QDs with two representative gallium ratio is shown in Fig. \ref{fig6}, which is compared with those of InP and GaP QDs of equal size. Two distinct characteristics are observed, (i) due to the lowering in symmetry, the degenerate excitonic absorption peaks of InP QDs are split with randomly incorporating Ga atoms into the lattice. The splitting is enhanced when increasing Ga ratio from $x=0.2$ to 0.6 (cf. Fig. \ref{fig6}(c, d)). (ii) The absorption spectrum blueshifts with increasing the Ga ratio. However, the absorption intensity, at least the lower energy part, decreases. Strikingly, we find that the absorption intensity for the first bright exciton state experiences a sudden drop to nearly zero at $x \approx 0.8$ (not shown), which might be related to switch in the band-edge transition from a $\Gamma_{5} \rightarrow \Gamma_{1}$ transition to a $\Gamma_{5} \rightarrow \Gamma_{5}$ transition. It should be pointed out that the direct gap to indirect gap transition takes place at $x \approx 0.77$ for bulk Ga$_{x}$In$_{1-x}$P\cite{Merle2032}.

\section{Conclusion}
\label{sec_col}
To summarize, we have presented a detailed study of structural, electronic and excitonic optical properties of InP, GaP and their ternary compound GaInP QDs of realistic sizes. The electronic structure is calculated using a hybrid functional within density functional theory, while the optical properties are accounted for based on the time-dependent density functional theory. We find that single-particle gap of InP QDs scales nearly linearly as a function of the inverse diameter. The excitonic effects have only a marginal impact on this scaling law, and the calculated optical gaps are found in excellent agreement with available experiments. The exciton binding energy scales as $1/D^{0.77}$, not as $1/D$ as expected, while the radiative exciton decay lifetime is found to increase surprisingly linearly as a function of dot size. For GaP QDs, we have predicted an electron state crossover whereby the nature of the lowest unoccupied molecular orbital (LUMO) state changes its symmetry from $\Gamma_{5}$ to $\Gamma_{1}$ at diameter around 1.5 nm. After the crossover, the pronounced band-edge exciton state is dominantly contributed from the $\Gamma_{5} \rightarrow \Gamma_{1}$ transition. Both the singlet-triplet splitting and the intensity of the lower energy part of the absorption spectrum experience a significant enhancement with increasing the quantum confinement effects.
Finally, we find that Vegard's law holds very well in the GaInP random alloyed quantum dots. The bowing parameter of this common-cation alloyed quantum dot appears size-dependent and much smaller than its bulk parentage. The physical mechanism responsible for the energy gap bowing is mainly ascribed to the volume deformation. The excitonic effects are found to have only marginal impact on the energy gap bowing. The current study could be helpful for gaining insight into the electronic and optical properties of colloidal quantum dots of group III-V towards future optoelectronic applications.

\section*{Acknowledgement}
The work has been partly supported by  NSFC project with grant No. 11804077 and  11774078, and partly by the innovation research team of science and technology in Henan province (20IRTSTHN020) and the Distinguished Professor grant of Henan University with grant No. 2018001T.

\providecommand{\noopsort}[1]{}\providecommand{\singleletter}[1]{#1}%

\end{document}